\documentclass[twocolumn,showpacs,preprintnumbers,amsmath,amssymb,prl]{revtex4}

\usepackage{graphics}
\usepackage{epsfig}
\usepackage{bm}

\def\beq{\begin{equation}}
\def\eeq{\end{equation}}

\def\reff#1{(\ref{#1})}

\def\subsc#1{{\mbox{\rm\scriptsize #1}}}

\def\rhocrit{\rho_\mathrm{c}}
\def\Up{U_\mathrm{p}}
\def\Ni{N_\mathrm{i}}

\def\Etot{E_\mathrm{tot}}

\def\Wcmcm{\mbox{\rm Wcm$^{-2}$}}
\def\abl#1#2{\frac{\mbox{\rm d} #1}{\mbox{\rm d} #2}}

\def\N3d{N_\subsc{3D}}

\def\omegaMie{\omega_\mathrm{Mie}}
\def\omegaplasma{\omega_\mathrm{p}}
\def\omegalaser{\omega_\mathrm{l}}

\def\vekt#1{\bm{#1}}

\def\vektr{\vekt{r}}

\def\vektE{\vekt{E}}

\def\vektnabla{\vekt{\nabla}}

\def\vektEsc{\vektE_\mathrm{sc}}

\def\Edach{E_0}

\def\Ehat{\Edach}

\def\diff{\,\mbox{\rm d}}

\begin{document}

\title{Nonlinear resonance absorption in laser-cluster interaction}
\date{\today}
\author{M.\ Kundu and D.\ Bauer}
\affiliation{Max-Planck-Institut f\"ur Kernphysik, Postfach 103980, 69029 Heidelberg, Germany}
\date{\today}

\begin{abstract}
Rare gas or metal clusters are known to absorb laser energy very efficiently. Upon cluster expansion the Mie plasma frequency may become equal to the laser frequency. This linear resonance has been well studied both experimentally and theoretically employing pump probe schemes.
In this work we focus on the few-cycle regime or the early stage of the cluster dynamics where linear resonance is not met but nevertheless efficient absorption of laser energy persists. By retrieving time-dependent oscillator frequencies from particle-in-cell simulation results, we show that nonlinear resonance is the dominant mechanism behind outer ionization and energy absorption in near infrared laser-driven clusters.
\end{abstract}


\maketitle

The construction of laser-based table-top sources of energetic electrons, ions, and photons requires the efficient coupling of the incident (usually $800$--$1064$\,nm) laser light with matter. Clusters turned out to be very efficient absorbers of laser light since they combine the advantages of gas and solid targets, namely transparency and high charge density.
Almost 100\% absorption of the laser light was observed in experiments with rare gas clusters \cite{ditm97}. 

The interaction scenario for laser wavelengths $\geq 800$\,nm, on which many researchers in the field agree upon, is as follows: after removal of the first electrons from their ``parent'' ions (inner ionization) and the cluster as a whole (outer ionization) the cluster charges up. The total electric field (i.e., laser plus space charge field) inside the cluster leads to inner ionization up to high charge states not possible with the laser field alone (ionization ignition \cite{rose97,bauer03}). However, the restoring force of the ions counteracts outer ionization so that ionization ignition stops at some point. Moreover, the cluster expands due to Coulomb explosion and thermal pressure, thus lowering the electric field due to the ions. The latter determines the dominant eigenfrequency of the cluster, i.e., the Mie frequency $\omegaMie(t)={\omegaplasma(t)}/{\sqrt{3}}=\sqrt{{4\pi \rho(t) }/{3}}=\sqrt{{\Ni(t) Z(t)}/{R^3(t)}}$
with $\rho$ the charge density, $\Ni$ the number of ions of (mean) charge state $Z$, and $R$ the cluster radius (atomic units are used unless noted otherwise). For laser wavelengths $\geq 500$\,nm, soon after the removal of the first electrons $\omegaMie$ exceeds the laser frequency $\omegalaser$. Hence linear resonance $\omegaMie(t)=\omegalaser$
occurs not before the cluster has sufficiently expanded (typically after a few hundred femtoseconds). At linear resonance the electric field inside the cluster is enhanced instead of shielded \cite{ditm96} so that even higher charge states can be created and even more energy can be absorbed from the laser. 

Absorption of laser energy is only possible through resonances (linear or nonlinear)  or non-adiabaticities (collisions). Commonly used phrases to explain absorption, like the tautological ``laser dephasing heating'' or the ill-defined ``collisions with the cluster boundary,'' are correct but meaningless since dephasing is, according to Poynting's theorem, a prerequisite for absorption while electron collisions with the ``cluster boundary'' occur all the time, independent of whether absorption is efficient or not.

     The importance of the linear resonance has been demonstrated both in pump probe experiments and simulations \cite{doepp05,koell99,zam04,last99,saal03,fenn04,sied05,mart05}. The emission of third and higher harmonics of the incident laser light has been observed in computer simulations \cite{fomi05}. Collisional ionization and absorption are of minor importance at wavelengths $\simeq 800$\,nm or greater \cite{ishi00,megi03,bauer04} whereas it is the dominant absorption mechanism at short wavelengths \cite{bauer04,sied04,laar05} not studied in the present Letter.

One of the crucial points in the above mentioned scenario is the mechanism of outer ionization, which goes hand in hand with absorption \cite{gresch05} since the laser energy is transiently stored in the freed, energetic electrons. These electrons leave net positive charge behind, which finally Coulomb explodes. The latter converts electron energy into ion energy, which explains why experimentalists typically measure MeV ions but only keV electrons \cite{ditmNature,kum03,spring03}. In order to separate outer ionization and absorption due to linear resonance from other mechanisms we consider only the first few tens of femtoseconds of the laser-cluster interaction where ion motion is negligible and linear resonance is therefore ruled out. Previous work pointed out already the possible importance of nonlinear resonance \cite{tagu04,mulser05} or, equivalently, Landau damping in finite systems \cite{korn05}. So far these mechanisms were clearly observed only in simplified model systems. The interpretation of results from molecular dynamics or particle-in-cell simulations, on the other hand, is often hampered by the complex dynamics of the individual particles that makes the clear distinction of absorption mechanisms difficult. In this Letter we bridge this gap by analyzing our particle-in-cell results in terms of nonlinear oscillators. This enables us to prove that essentially {\em all} electrons contributing to outer ionization pass through the nonlinear resonance, so that the latter is unequivocally identified as {\em the} collisionless absorption mechanism if linear resonance is impossible.

In general, the eigenfrequency of a particle in a given potential depends on the excursion amplitude (or the energy) of the particle, $\omega=\omega[\hat{\vektr}]$. In a (laser-) driven system the excursion amplitude is time-dependent so that it may dynamically pass through the nonlinear resonance (NLR) \beq \omega[\hat{\vektr}(t)] = \omegalaser.\eeq
Refs.\ \cite{mulser05,baumu05} discuss at length the significant features of laser-driven systems undergoing NLR. Here we restrict ourselves to the prerequisites necessary to understand the analysis of our particle-in-cell results below.

\begin{figure}
\includegraphics[width=0.48\textwidth]{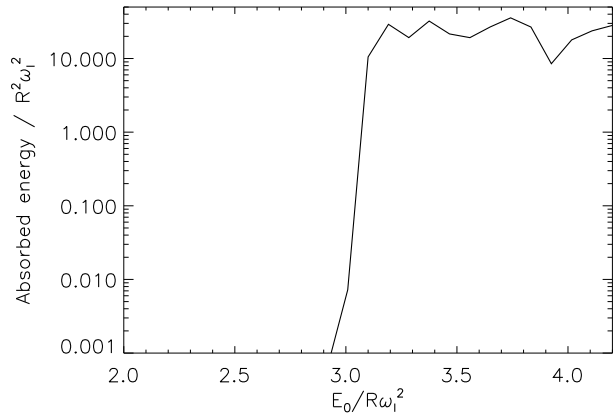}
\caption{Laser absorption vs.\  driver amplitude in the rigid sphere-model \reff{rsm_eom} for $\omegaMie/\omegalaser=2.7$, $E(\tau)=-\Ehat\sin^2(\tau/2n)\cos(\tau)$, $n=10$. The nonlinear resonance is passed once the threshold driver amplitude $\simeq 3.0$ is reached. \label{rsmI}}
\end{figure}

\begin{figure}
\includegraphics[width=0.48\textwidth]{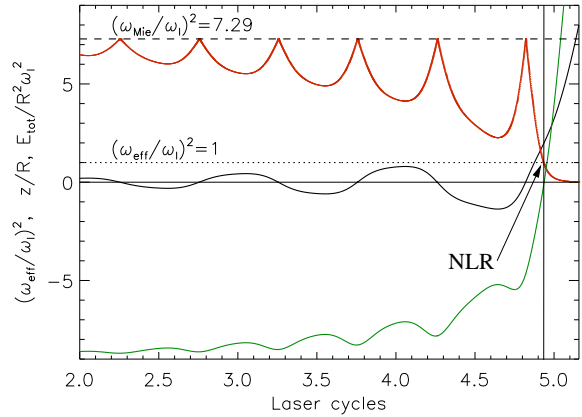}
\caption{(color online). Typical behavior of $\left({\omega_\mathrm{eff}(\tau)}/{\omegalaser}\right)^2$ (drawn red) vs.\  time above the threshold driver strength in Fig.~\ref{rsmI} (actual value was $\Ehat/R\omegalaser^2=3.3$). Excursion $z/R$ (black) and  energy of the electron sphere $\Etot/R^2\omegalaser^2$ (green) are included in the plot. Outer ionization (i.e., $\Etot/R^2\omegalaser^2\geq 0$) and occurrence of nonlinear resonance $ \left(\omega_\mathrm{eff}(\tau)/\omegalaser\right)^2=1$ (dashed-dotted line) always coincide (vertical line).  \label{rsmII}}
\end{figure}

In the rigid sphere-model (RSM) (see, e.g., \cite{parks01,fomi03,mulser05}) of a cluster both electrons and ions are modelled by homogeneously charged spheres which, in a linearly polarized laser field, oscillate along $z$ around their common center of mass. In the case of equal charge density and radii $R$, the equation of motion for the center of the electron sphere can be written in dimensionless entities as
\beq \abl{^2r}{\tau^2} + \left( \frac{\omegaMie}{\omegalaser}\right)^2 \mathrm{sgn}(z) \left\{ \begin{array}{c}\displaystyle r-\frac{9r^2}{16}+\frac{r^4}{32} \\ \displaystyle r^{-2} \end{array}\right\} = \frac{E(\tau)}{R\omegalaser^2}. \label{rsm_eom}\eeq
Here, $r=\vert z\vert /R$, $\tau=\omegalaser t$, and the upper line applies for $0\leq r <2$, the lower line (Coulomb force) for $r\geq 2$. 
As was observed in Ref.\ \cite{mulser05} the absorption of laser energy in the RSM is characterized by a threshold driver strength below which absorption is negligible (harmonic regime) and above which absorption is almost constant. Figure \ref{rsmI} shows this threshold behavior for $\omegaMie/\omegalaser=2.7$ and a $n=10$-cycle $\sin^2$-pulse $E(\tau)=-\Ehat\sin^2(\tau/2n)\cos(\tau)$. 
Equation \reff{rsm_eom} can be formally rewritten as $\diff^2 r/\diff\tau^2+\left({\omega_\mathrm{eff}[\hat{r}(\tau)]}/{\omegalaser}\right)^2 \mathrm{sgn}(z)\, r = {E(\tau)}/{(R\omegalaser^2)}$ with
\beq \left(\frac{\omega_\mathrm{eff}(\tau)}{\omegalaser}\right)^2= \frac{1}{\mathrm{sgn}(z)(\tau)\, r(\tau)}\left( \frac{E(\tau)}{R\omegalaser^2} - \ddot{r}(\tau) \right) \label{efffrequ} \eeq
the instantaneous, scaled effective frequency ${\omega_\mathrm{eff}(\tau)}/{\omegalaser}$, which passes through unity at the NLR.
Figure \ref{rsmII} shows a typical example for the temporal behavior of $ \left(\omega_\mathrm{eff}(\tau)/\omegalaser\right)^2$ above the threshold driver strength in Fig.~\ref{rsmI}. Since $(\omegaMie/\omegalaser)^2=7.29$, $ \left(\omega_\mathrm{eff}(\tau)/\omegalaser\right)^2$ starts at this value (dashed line in Fig.~\ref{rsmII}) and drops with increasing driver strength.  It passes through unity at the time indicated by the vertical line, and it is exactly at that time where the electron sphere is set free, as it is clearly visible from the energy of the electron sphere, which passes through zero, and the excursion. We have checked numerically that NLR occurrs for all driver strengths above the threshold whereas the resonance is never met below threshold. In rare cases above the threshold, the electron sphere returns to the origin and recombines. In general, however, NLR results in a detached electron sphere at the end of the laser pulse.

\begin{figure}
\includegraphics[width=0.51\textwidth]{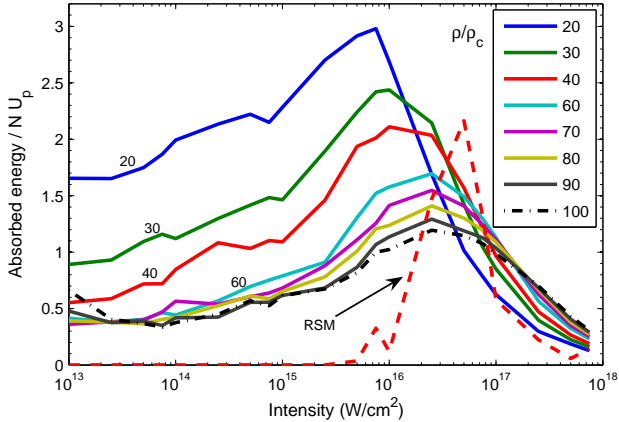}
\caption{(color online). PIC results for a Xe$_\mathrm{1600}$ cluster ($R=3.2$\,nm). Total absorbed energy per electron in units of $\Up$ vs.\  laser intensity for charge densities between $20$ and $100$ times the critical density, corresponding to average charge states between $1.6$ and $8$. The prediction of the RSM  for $\rho/\rhocrit=40$ is included in the plot (dashed).  \label{PICabsenerg}}
\end{figure}

Let us now turn to the particle-in-cell (PIC) \cite{birdsall} results. We consider pre-ionized clusters of fixed radius $R=3.2$\,nm (e.g., Xe$_{N_\mathrm{i}}$ with $N_\mathrm{i}\simeq 1600$) but of various charge densities (i.e., different degree of inner ionization). The ratio of charge density to critical density $\rho/\rhocrit=3\omegaMie^2/\omegalaser^2$ varies from $20$ to $100$. The clusters are exposed to $8$-cycle $\sin^2$-pulses of wavelength $\lambda=1056$\,nm. Since ion motion does not play an important role during the simulation time, the ions are fixed, which ensures a well defined, constant Mie frequency $\omegaMie$. Electron-ion collisions are neglected in our PIC treatment so that absorption of laser energy can only proceed through other non-adiabaticities (like collisions with the ``cluster boundary'' or the cluster as a whole) or resonances. Figure \ref{PICabsenerg} shows the absorbed energy per electron in units of the ponderomotive potential $\Up=\Ehat^2/4\omegalaser^2$, i.e., the time-averaged quiver energy of a free electron in the laser field. 
One sees that the absorbed energy per electron is always on the order of $\Up$. However, the absorbed energy is nonlinear in $\Up$ and displays a maximum before it drops because outer ionization saturates. The maxima are located close to the threshold intensities predicted by the RSM (see RSM-result for $\rho/\rhocrit=40$). 
With increasing charge density the maxima of the absorbed energy (divided by $\Up$)  move towards higher laser intensities while the absorbed energy per electron decreases.  

The motion of the PIC particles can be analyzed in the same way as it was done with the motion of the electron sphere in the RSM above. A PIC particle has the same charge to mass ratio as a ``real'' electron, that is, $e/m=-1$ in atomic units. Each PIC particle moves under the influence of the external laser field and the space charge field  $\vektEsc=-\vektnabla \Phi(\vektr,t)$ due to the potential $\Phi(\vektr,t)$ that is created by all charges (mapped to a numerical grid). Hence the equation of motion of the $i$th PIC particle is $ \ddot{\vektr}_i + \vektEsc(\vektr_i,t) = -\vektE(t)$. 
The equation for the effective, time-dependent oscillator frequency analogous to \reff{efffrequ} then reads
\beq \omega^2_{\mathrm{eff},i}(t) = -\frac{[\vektE(t)+\ddot{\vektr}_i]\cdot\vektr_i}{\vektr^2_i}=\frac{\vektEsc(\vektr_i,t)\cdot\vektr_i}{\vektr^2_i} \label{efffrequPIC}.\eeq
Clearly, $\vektEsc(\vektr_i,t)$ depends on the position of {\em all} other particles $\neq i$ as well. The PIC simulation starts with the neutral cluster configuration, i.e., the positive charge by the ionic background is neutralized by the negative charge of the PIC particles, so that $\vektEsc(\vektr_i,0)\equiv \vekt{0}$. Hence, a PIC electron  ``sees'' initially an effective frequency $\omega_{\mathrm{eff},i}(0)=0$. The laser field disturbs the charge equilibrium and $\omega^2_{\mathrm{eff},i}(t)$ becomes different from zero. $\omega^2_{\mathrm{eff},i}(t)$ may be even negative in regions of accumulated electron density (repulsive potential). As the cluster charges up, $(\omega_{\mathrm{eff}}/\omegalaser)^2$ quickly increases beyond unity (where the RSM starts in the first place). In the PIC simulation each electron experiences its own  $\omega_{\mathrm{eff},i}$, depending on the position. Hence, a sharp threshold in absorption as in Fig.~\ref{rsmI} does not exist. Besides this, the starting from $\omega_{\mathrm{eff},i}(0)=0$, the possibility of negative $\omega^2_{\mathrm{eff},i}(t)$, and the three-dimensionality are the main differences to the RSM analysis above.

\begin{figure}
\includegraphics[width=0.485\textwidth]{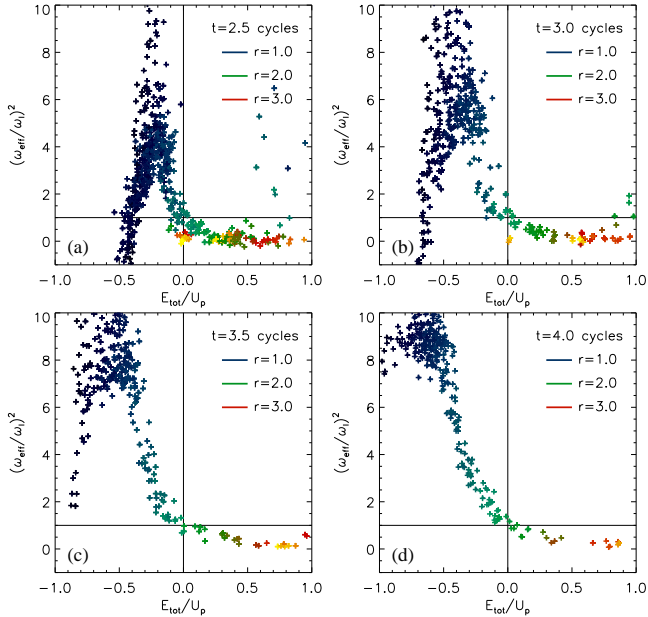}
\caption{(color online). Snapshot of PIC electrons in the frequency vs.\  energy-plane at times (a) $t=2.5$, (b) $t=3.0$, (c) $t=3.5$, and (d) $t=4.0$ laser cycles. The laser intensity was $2.5\times 10^{16}$\,\Wcmcm and $(\omegaMie/\omegalaser)^2=40/3$. Other parameters as in Fig.~\ref{PICabsenerg}. The radial positions (in units of $R$) are color-coded. Electrons become free upon crossing the nonlinear resonance, i.e., $(\omega^2_{\mathrm{eff}}/\omegalaser^2,\Etot/\Up)=(1,0)$. At that time the radial position is $r\simeq 2$.   \label{PICweffvsE}}
\end{figure}

By following the dynamics of the electrons in the effective frequency vs.\  energy-plane one can identify the main pathway to outer ionization and efficient absorption. In Fig.~\ref{PICweffvsE}a--d the scaled effective frequencies squared $(\omega_{\mathrm{eff}}/\omegalaser)^2$ of the individual PIC electrons are plotted vs.\  their energies $E_{\mathrm{tot},i}(t)=\dot{\vektr}_i^2(t)/2-\Phi(\vektr_i,t)$ they would have if the driver is switched off instantaneously at $t=2.5$, $3$, $3.5$, and $4$ laser cycles, respectively. We define the time when, for a particular electron, $\Etot$ becomes $>0$ as the ionization time of that electron. The laser intensity is $2.5\times 10^{16}$\,\Wcmcm, and the preionized cluster is $40$ times overcritical so that $(\omegaMie/\omegalaser)^2=40/3$. As is clearly visible in Fig.~\ref{PICweffvsE},  each electron reaches positive energy close to the point $(\omega^2_{\mathrm{eff}}/\omegalaser^2,\Etot/\Up)=(1,0)$. The radial position of each electron is color coded, indicating that outer ionization occurs at radii around $2R$. Data points at radii $>3R$ (orange and yellow colors in  Fig.~\ref{PICweffvsE}a,b) with positive but very small $\Etot$ and $\omega^2_{\mathrm{eff}}\simeq 0$ represent low energetic electrons removed earlier during the pulse. Electrons with positive energy but small radii [visible in (a) and (b)] are those driven back to the cluster by the laser field. For the electrons inside the cluster potential (negative energies and radii $<R$, color coded blue and black)  $(\omega_{\mathrm{eff}}/\omegalaser)^2$ spreads over a wide range, starting from the maximum value $(\omegaMie/\omegalaser)^2$ down to negative values due to the repulsive force exerted by the compressed electron cloud. Note that negative values occur mainly at early times where most of the electrons are still inside the cluster. The occurrence of NLR is less clear for the few electrons leaving the cluster even earlier than $t=2.5$ laser cycles. As mentioned above, these electrons move in a shallow effective potential with  $(\omega_{\mathrm{eff}}/\omegalaser)^2<1$ when they leave the cluster with ease and with rather low kinetic energy because the laser intensity is still low at the time of their emission. With increasing degree of outer ionization, the resonance point $(1,0)$ is met more and more precisely [i.e., less vertical spread around $(1,0)$] since the effective frequency is more and more dominated by the uniform, static ion background. The electrons pass very quickly through the resonance  (typically less than a quarter of a laser cycle) because the energy gain necessarily drives the electrons out of resonance again.

One may object that, since the denominator in \reff{efffrequPIC} necessarily increases while the numerator decreases for an electron on its way out of the cluster potential, the passage through a point $(\omega^2_{\mathrm{eff}}/\omegalaser^2,\Etot/\Up)=(x,0)$ with $x$ {\em some} value $< (\omegaMie/\omegalaser)^2$ is rather the consequence of outer ionization than the mechanism behind it. However, NLR only occurs at $x=1$, and the results in  Fig.~\ref{PICweffvsE} show only  little spreading along $(\omega_{\mathrm{eff}}/\omegalaser)^2$ at $\Etot=0$. Moreover, the fact that {\em both} the single electron energies become positive {\em and}  the radii exceed $\simeq 2 R$ when $(\omega_{\mathrm{eff}}/\omegalaser)^2= 1$ indicates that NLR is indeed the responsible mechanism behind outer ionization accompanied by efficient absorption of laser energy. One may further object that such a short resonance is more like a collision (e.g., with the cluster boundary) than a resonance. However, note that collisions with the cluster boundary occur many times but efficient absorption and outer ionization only takes place when the resonance condition is fulfilled.   

In summary, we have shown that cluster electrons contributing to efficient absorption and outer ionization in near infrared laser fields undergo nonlinear resonance, meaning that the instantaneous frequency of their motion in a time-dependent, anharmonic, effective potential meets the laser frequency. Nonlinear resonance is the only possible absorption mechanism if the laser pulse is too short for the linear resonance to occur (or during the early cluster dynamics in longer pulses) and if electron-ion collisions (inverse bremsstrahlung) are negligible, as it is the case at near infrared or greater wavelengths.   In order to prove the occurrence of nonlinear resonance we introduced a method to analyze the results obtained from particle-in-cell simulations, namely the mapping of the system of electrons and ions that interact through their mean field  onto a system of nonlinear oscillators whose time-dependent frequencies unequivocally revealed the coincidence of electron removal and nonlinear resonance.

Fruitful discussions with P.\ Mulser are gratefully acknowledged.
 This work was supported by the Deutsche Forschungsgemeinschaft.


\end{document}